\documentclass[twocolumn]{revtex4-2}

\usepackage{graphicx,epsfig}
\usepackage{amsmath}
\usepackage{amsfonts}
\usepackage{float}
\usepackage{fancyhdr}
\usepackage{booktabs}
\usepackage[table]{xcolor}
\usepackage{soul}
\usepackage{xr}
\usepackage{multirow}
\usepackage{xcolor}

\usepackage{ulem}
\begin{document}
\title{Continuous-time random walk model for the diffusive motion of helicases}

\author{V. Rodríguez-Franco$^{1}$, M.M. Spiering$^{2}$, F. Ritort$^{1,3,4}$, M. Mañosas$^{1,3}$}

\affiliation{$^{1}$ Small Biosystems Lab, Condensed Matter Physics Department, Universitat de Barcelona, Carrer de Martí i Franquès 1, 08028 Barcelona, Spain}

\affiliation{$^{2}$ Department of Chemistry, The Pennsylvania State University, University Park, PA, USA}

\affiliation{$^{3}$ Institut de Nanociència i Nanotecnologia, Universitat de Barcelona, 08029 Barcelona, Spain}

\affiliation{$^{4}$Reial Acadèmia de Ciències i Arts de Barcelona (RACAB), Secció Física.2a, La Rambla 115, E-08002 Barcelona, Spain}

\begin{abstract}
{\bf Abstract: } 

DNA helicases are molecular motors that use the energy from ATP hydrolysis to move along DNA, promoting the unwinding or rewinding of the double helix. Here, we use magnetic and optical tweezers to track the motion of three helicases, gp41, RecQ, and RecG, while they unwind or rewind a DNA hairpin. Their activity is characterized by measuring the helicase rate and diffusivity under different force and ATP conditions.   
We use a Continuous-Time Random Walk framework that allows us to compute the mean helicase displacement and its fluctuations analytically. Fitting the model to the measured helicase velocity and diffusivity allows us to determine the states and transitions in the helicase mechanochemical cycle.
A general feature for all helicases is the need to incorporate an off-pathway pausing state to reproduce the data, raising the question of whether pauses play a regulatory role. Finally, we determine the thermodynamic uncertainty factor and the motor efficiency, finding that the rewinding RecG helicase reaches high efficiencies that approach 100$\%$ when operating uphill. Incorporating the analysis of fluctuations allows to better characterize the activity of molecular machines, which represents an advance in the field.


\end{abstract}

\maketitle


\section{Introduction}

Helicases are ubiquitous enzymes that are present in all living organisms. They work as molecular motors, converting the chemical energy from the hydrolysis of adenosine triphosphate (ATP) into mechanical work and motion along nucleic acids (NA) \cite{lohman1996mechanisms,tuteja2004unraveling}. 
Helicase motion is coupled to various functions, such as: NA unwinding and separation of the two strands of the duplex (e.g. in DNA replication and transcription); NA rewinding and formation of multi-branch structures (e.g. in DNA recombination); and disruption of protein-NA interactions (e.g. in DNA repair and RNA processing) \cite{tuteja2004unraveling,delagoutte2003helicase,pyle2008translocation}.
These diverse functions are accomplished by the presence of different helicases, classified into families based on their conserved structural motifs 
\cite{caruthers2002helicase,singleton2007structure}.
Besides their relevance in molecular biology, helicases, and molecular motors in general, have raised interest in many other fields, and in physics, they are a paradigm of small systems \cite{bustamante2005nonequilibrium}.
Molecular motors typically move in nanometer (nm) steps while generating mechanical forces on the picoNewton (pN) range, leading to the generation of work of pN·nm, which is on the order of the thermal energy unit $k_B T$ ($1k_B T=4$pN·nm). On the other hand, the energy released by the hydrolysis of ATP is $\sim 10 k_B T$, which implies that these motors work on a strong Brownian environment where fluctuations play a central role.

The helicase translocation along the NA strand in a defined directionality (3' to 5' or 5' to 3') results from different nucleotide-enzyme conformations of distinct NA affinities connected in a cyclic reaction network \cite{delagoutte2002helicase,patel2006mechanisms}.
The mechano-chemical coupling between the ATP hydrolysis reaction and the enzyme motion can be described in terms of two mechanisms: the Brownian Ratchet (BR) and the Power Stroke (PS) \cite{parrondo1996criticism,patel2006mechanisms,pyle2008translocation,hwang2019structural,galburt2017conformational}.
These two mechanisms exemplify the two limit cases: biasing or rectifying the motor diffusive thermal motion (BR) versus downhill dynamics based on a motor structural change induced by ATP hydrolysis (PS). In practice, many molecular motors may employ a combination of both mechanisms to achieve efficient motion \cite{hwang2019structural}.
These two mechanisms can also be used to describe motor translocation coupled to NA strand separation/annealing during NA unwinding/rewinding, leading to the classification of helicases in active (for PS) and passive (for BR) \cite{galburt2017conformational,patel2006mechanisms,manosas2010active}.
The active helicase directly interacts with the ssNA/dsNA junction, destabilizing (or stabilizing) the base-pairs (bp) at the fork before translocation. In contrast, the passive helicase relies on the thermal fraying of the bp at the NA fork to promote un(re)winding. 


Traditionally, helicase activity has been characterized using ensemble assays, such as gel-based, fluorescence, or spectrofluorometry \cite{matson1983gene,raney1994fluorescence,kim2009vitro}. These approaches measure average properties over a large ensemble of molecules, but they provide limited information about nanoscale processes where fluctuations are relevant. In the last 30 years, different single molecule techniques have emerged \cite{neuman2008single,joo2008advances}, allowing to monitor the activity of individual enzymes in real-time. These measurements facilitate the detection of molecular heterogeneity, rare events, pathways, intermediates, and dynamical NA-helicase interactions \cite{yodh2010insight}, crucial aspects for understanding how helicases work.
The movement of single helicases can be monitored using single-molecule force spectroscopy techniques, such as optical, magnetic or nanopore tweezers \cite{neuman2008single,craig2017revealing}. In these assays, a mechanical force is applied on the enzyme or along the NA substrate, and the NA extension along the force direction is measured, defining a reaction coordinate to follow the advance of the helicase \cite{tinoco2002effect}. By modifying the nucleotide conditions (e.g. ATP and ADP concentrations)
and the applied force, the coupling between the helicase motion and the ATP hydrolysis reaction can be investigated, discriminating between different helicase mechanisms
\cite{manosas2010active,dumont2006rna,ribeck2010dnab,spies2014two,seol2019homology,laszlo2022sequence}. 
The general approach has been to use the experimental data to test specific models for each helicase, finding that, in many cases, different helicases display distinct activities involving complex reactions with multiple kinetic pathways and/or different rate-limiting steps.

Most descriptions of molecular motors are based on simple kinetic models and pathways. From muscle transport motors such as kinesin and myosin, to genomic maintenance machines such as polymerases and helicases, most models have focused on describing the average motor speed under different conditions of ATP, ADP, force and temperature \cite{kolomeisky2007molecular,astumian2010thermodynamics,keller2000mechanochemistry}.
However, a main feature of these machines is the Brownian fluctuations and their diffusivity. Although diffusivity measurements are scarce, they provide clues about the motor mechanism and its efficiency. Despite its importance, helicases diffusive behaviour has not been analysed in detail before. The diffusivity ($D$), along with the velocity ($v$), are of particular interest as they are related to the randomness parameter $r=\frac{D}{dv}$, where $d$ is the motor step-size \cite{svoboda1994fluctuation}.
This parameter provides information about the number of rate-limiting steps in an enzymatic cycle, being $r=1$ for a Poisson process and $r=0$ for a molecular clock. Alternatively, the motor step size $d$ can be estimated if the number of rate-limiting steps is known
\cite{neuman2005statistical}. Additionally, $v$ and $D$ are related to the $Q$ factor of the thermodynamic uncertainty relation (TUR) \cite{barato2015thermodynamic,song2021thermodynamic}. The TUR sets an inequality between the entropy production rate $\sigma$ and the measurement precision in nonequilibrium steady states. It is defined in terms of generic nonequilibrium currents, which for the case of a translocating motor takes the form $Q=\sigma\frac{2D}{k_{B}v^{2}}\ge 2$ with $\sigma$ expressed in $k_B$ units \cite{song2021thermodynamic}. This factor quantifies the irreversibility of an enzymatic process in the non-linear regime and is related to the motor's thermodynamic efficiency $\eta$ defined as the ratio between the amount of delivered mechanical work ($W$) and the input chemical energy coming from the ATP hydrolysis ($\Delta \mu$): $\eta=\frac{W}{\Delta \mu}$. The two quantities satisfy the energy balance relation, $\Delta \mu=W+T\sigma t$, where $T\sigma$ stands for the heat's rate released to the environment. The thermodynamic efficiency $\eta$ can be expressed in terms of $Q$, $\eta=(1+Q\frac{vd}{2D}\frac{k_{B}T}{W})^{-1}$ \cite{song2021thermodynamic,pietzonka2016universal}, being $T$ the temperature. Small values of $Q$ indicates a more efficient motor that operates closer to the limits of thermodynamic optimization.

In this work, we have investigated three different DNA helicases using magnetic tweezers (MT) and optical tweezers (OT). Two of them catalyse DNA unwinding: the T4 gp41 helicase, which is involved in DNA replication in T4 bacteriophage, and RecQ from Escherichia Coli ($E.Coli$), which participates in different DNA repair pathways. The third one is the RecG helicase from $E.Coli$, which is involved in DNA repair and recombination, catalysing DNA annealing and the formation of multi-branched DNA structures \cite{venkatesan1982bacteriophage,manosas2010active,manosas2013recg,lionnet2007real,bagchi2018single,umezu1990escherichia,mcglynn2001rescue}.
In the experiments, a constant force is applied to the extremities of a DNA hairpin, and the helicase un(re)winding activity is followed by measuring the changes in the DNA extension, enabling real-time monitoring of the enzyme activity. 
From these measurements, we can infer the helicase position along the DNA and extract its mean rate, diffusivity, and pause kinetics. To interpret the experimental results, we have developed a general theoretical framework for helicase motion based on random walk theory, allowing us to compute the average helicase rate and diffusivity analytically. The model features a continuous-time random walk (CTRW) in a one-dimensional chain with an auxiliary pausing state. There are three distinct kinetic transitions: forward, backwards and enter/exit a pause. Each transition is described by activated kinetic rates that depend on force and ATP concentration ([ATP]). The model fits the translocation and unwinding/rewinding rates and diffusivity data for the different helicases over a wide range of forces and [ATP]. The fitting procedure allows us to determine the chemical (ATP concentration) and mechanical (force) dependencies of the kinetic rates connecting the different states, providing insights into the helicase mechano-chemical cycle. Assuming a tight mechano-chemical coupling, we have also investigated how the $Q$ factor and the motor efficiency $\eta$ changes for each helicase and how their values depend on the helicase step size and its active and passive nature.

\section{Methods}

\subsection{DNA substrates}

Two different hairpins, h1.2 and h1.4, of $\sim$1.2 and $\sim$1.4 kbp stems respectively, have been used for the MT and OT assays. Both hairpins have a 5' single-stranded DNA (ssDNA) tail of $\sim$80 nucleotides (nts) labelled with a biotin and 3' ssDNA tail of $\sim$100 nts labelled with several digoxigenins. 
The h1.2 hairpin was prepared as previously described in \cite{manosas2009coupling}. The stem of the h1.4 hairpin comes from a segment of the plasmid PBR322 and four different oligonucleotides are annealed and ligated to generate the loop and handles (details in Supplementary Fig. S1). Experiments with $E.Coli$ RecQ were performed with the h1.2 hairpin and experiments with RecG and gp41 were performed with the h1.4 hairpin.

\subsection{Enzyme preparation and experimental conditions}
The different helicases, T4 gp41, $E.Coli$ RecQ and $E.Coli$ RecG, were purified as previously described \cite{valentine2001zinc, bernstein2003domain, slocum2007characterization}. 


Assays with gp41 and RecG were performed in a buffer containing 25 mM Tris–Ac (PH 7.5), 10 mM $\rm {Mg(OAc)}_{2}$, 150 mM KOAc, 1 mM dithiothreitol (DTT), and different ATP concentrations (0.5-4 mM for gp41 and 100 $\mu$M-2 mM for RecG). 
Assays with RecQ were performed in a buffer containing 20 mM Tris–HCl (pH 7.5), 25 mM NaCl, 3 mM $\rm {Mg(Cl)}_2$, 1 mM DTT and different ATP concentrations (40 $\mu$M to 1 mM). All experiments were done at 25°C. The protein concentration was 50 nM for gp41 (monomeric concentration), 10 nM for RecG and 30 pM for RecQ.

\subsection{Single-molecule experiments}\label{Single-molecule experiments}

\begin{figure*}[htbp]
 \centering
 \includegraphics[width=\textwidth]{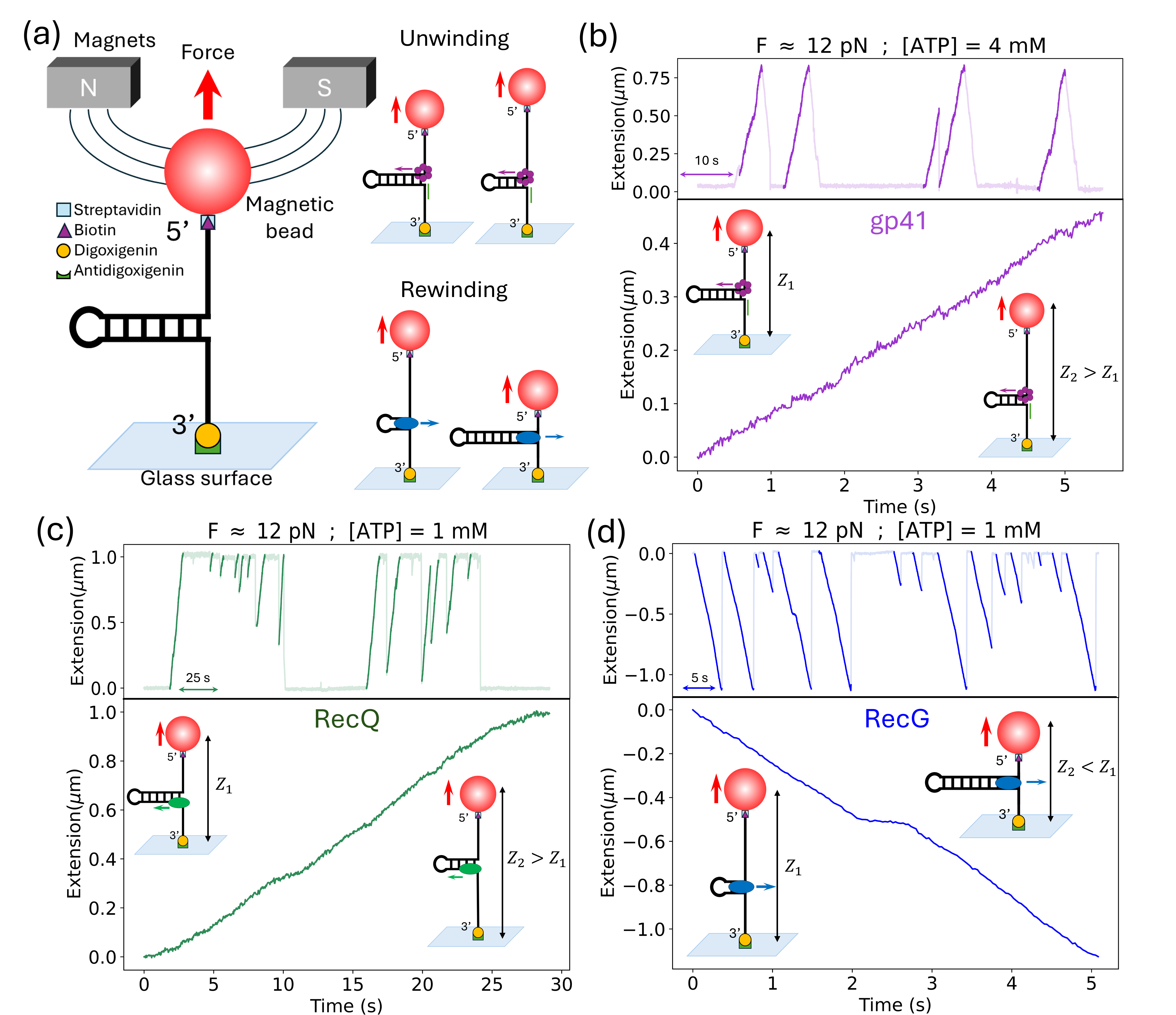}
 \caption{ \textbf{Helicase experiments} \textbf{(a)} Schematic representation of the MT experimental setup, where a DNA hairpin is tethered between a glass surface and a magnetic bead. 
 The progress of the unwinding and rewinding reactions leads to changes in the molecule extension.  \textbf{(b,c,d)} Experimental traces showing the hairpin extension as a function of time for gp41 (purple), RecQ (green) and RecG (blue) helicases. The un(re)winding events are highlighted in dark colour. The bottom panels show the details of a single un(re)winding trace with the schematic representations of the hairpin state at the beginning (left) and at the middle (right) of the trace.}
 \label{fig:traces and set up}
\end{figure*}

In magnetic tweezers (MT) experiments, we use a PicoTwist MT instrument (www.picotwist.com) to manipulate DNA hairpins tethered between a micrometric magnetic bead and the glass surface of a microfluidic chamber (Fig. \ref{fig:traces and set up} (a)). For making the tethers the glass surface is treated with an anti-digoxigenin antibody and passivated with bovine serum albumin
and the micron-sized magnetic beads are coated with streptavidin (Invitrogen MyOne). The applied force is controlled by adjusting the distance between the magnets and the sample ($Z_{mag}$). The microfluidic chamber is illuminated by a red LED that generates a parallel and monochromatic illumination. Using an inverted microscope connected to a CMOS camera we image the beads. The images are decorated by a set of diffraction rings, enabling real-time tracking of beads' 3D position with nanometric resolution at 30-80 Hz \cite{gosse2002magnetic,lionnet2012single}. From the bead's $z$ position we obtain the extension of the DNA molecule. The bead's fluctuations in the $x-y$ plane are used to measure the force via the equipartition theorem \cite{gosse2002magnetic,lionnet2012single}. An average calibration curve $F(Z_{mag})$ is used to estimate the force with 10$\%$ error due to bead inhomogeneities \cite{lionnet2012single}. 
In OT experiments, the DNA molecule is tethered between two micron-sized polystyrene beads using biotin-streptavidin and digoxigenin-antidigoxinein bonds. One bead is immobilized on the tip of a micropipette and the second bead is captured in an optical trap generated by two counter-propagating lasers (Supplementary Fig. S4 (a)). The force acting on the bead can be measured from the change in light momentum deflected by the bead using position-sensitive detectors \cite{smith20037}. After injecting the helicases and ATP, unwinding and rewinding activities are detected as an increase and decrease of the measured DNA hairpin extension respectively (see Fig. \ref{fig:traces and set up} and Supplementary Fig. S4). 

MT are used to test the activity of helicases in a force range going from 5 to 15 pN. Typically about 50-100 beads are tracked simultaneously which allows us to obtain large statistics. 
For the unwinding helicases (gp41 and RecQ), we started with the hairpin formed at a force below the unzipping force (15 pN) and monitored the unwinding activity from the DNA extension changes (Fig. \ref{fig:traces and set up} b,c). The gp41 helicase requires a long 5' tail for efficient loading \cite{valentine2001zinc,feng2023structural}. For this reason, we used a 50-mer oligonucleotide that is complementary to a hairpin region located at $\sim$500 bp from the fork. By unzipping the hairpin (at $F>15$ pN) we hybridized the oligonucleotide, generating a $\sim$500 nt 5' and 3' tails (see Supplementary Fig. S2). 
For the RecG rewinding helicase, we initially started with a partially unzipped hairpin. The partially unzipped configuration was achieved by either hybridizing a short oligonucleotide complementary to a hairpin region or by using forces close to the unzipping force. In the former case the force was first increased to mechanically unzip the hairpin and allow the oligonucleotide to bind. Next, the force was decreased to a given value (typically between 5 and 13 pN) and the hairpin partially reforms until reaching the oligonucleotide that blocks the full re-zipping (Supplementary Fig. S3 (a)). In the later case, the force was set to a value ($\sim 15$ pN) where the hairpin unzips except the last $\sim$50 bases containing a GC-rich region that require larger forces to unzip (see Supplementary Fig. S3 (b)). In both cases, the rewinding of the partially unzipped hairpin was detected as a decrease in the measured DNA extension (see Fig. \ref{fig:traces and set up} and Results). 
OT were also used to test the activity of the RecG helicases at larger forces (above 15 pN). In OT assays, we initially increased the distance between the micropipette and the trap ($X_{T}$) reaching a force of $\sim$15 pN and a partially unzipped hairpin. In the presence of RecG, the rewinding reaction causes a shortening of the DNA that induces the displacement of the bead in the trap, generating an increase in force (Supplementary Fig. S4 (b)). By using a force feedback protocol we can test the rewinding activity at different forces. We did not observe any activity above $\sim$35 pN, in agreement with previous measurements \cite{manosas2013recg}. These large forces probably induce the stalling and dissociation of the enzyme from the DNA template.

\subsection{CTRW Model} \label{CTRW model}
\begin{figure*}[htbp]
 \centering
 \includegraphics[width=\textwidth]{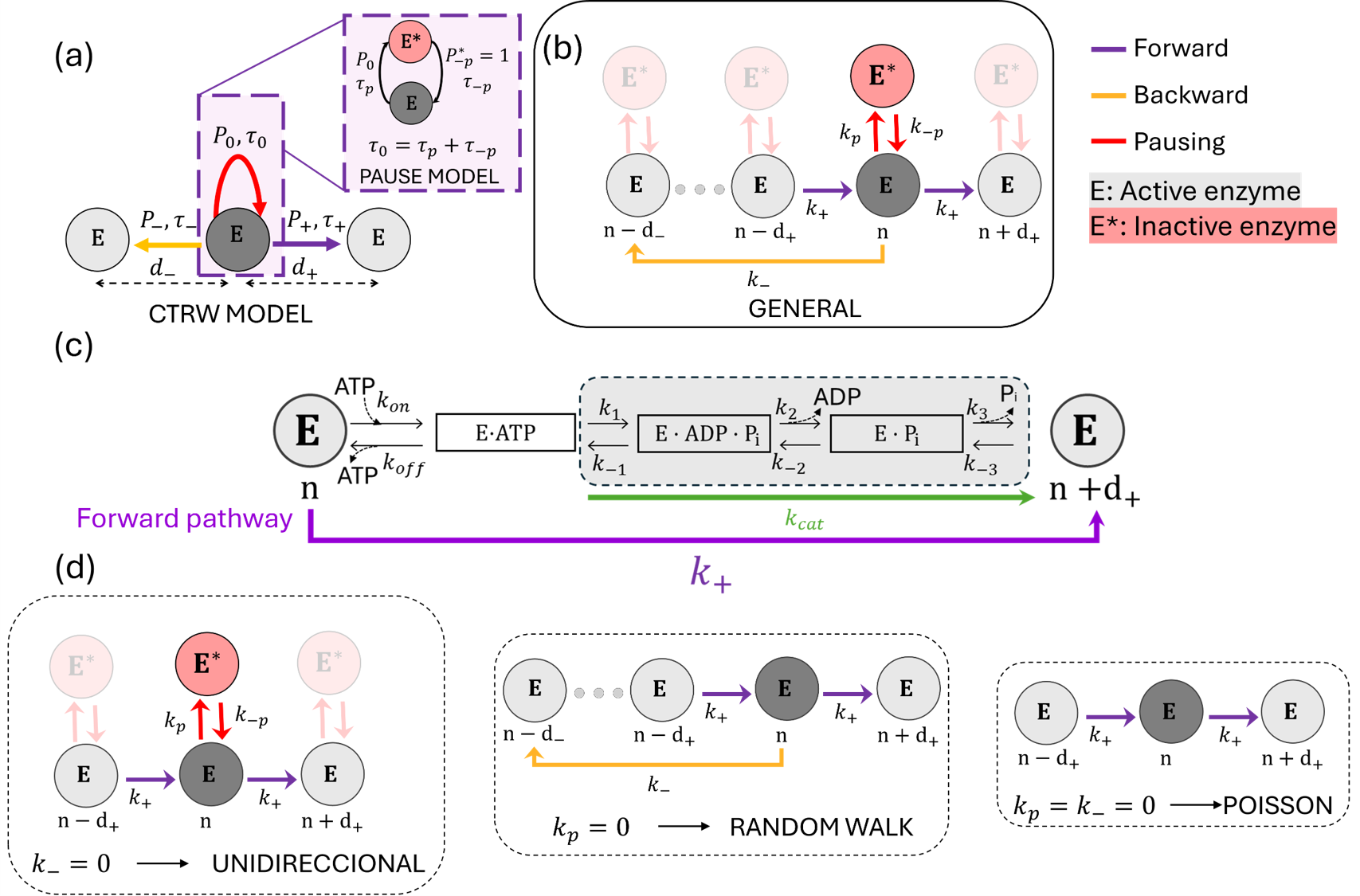}
 \caption{ {\bf Helicase models. }\textbf{(a)} Diagram of the CTRW model used to describe helicase motion. Inset shows the diagram of the pause transition divided into two steps corresponding to entering and exiting the pause state. $E$ and $E^{*}$ represent the states of a translocation-active and a pause-inactive helicase \textbf{(b)} Diagram of the general model with kinetic rates, which includes three pathways: the forward ATP-hydrolysis coupled to translocation on-pathway transition (purple), the off-pathway backward transition (yellow) and the off-pathway pausing transition (red). Kinetic rates and distances between states are indicated for each transition. \textbf{(c)} Details 
 of the forward ATP-hydrolysis coupled to the translocation pathway indicating the intermediate steps. The overall rate $k_{+}$ (purple) integrates the ATP binding and unbinding reaction as well as the ATP hydrolysis and release of ADP and Pi. The ATP hydrolysis reaction (grey shaded area) is highly irreversible and can be approximated with a single kinetic rate $k_{cat}$ (green). \textbf{(d)} Sub-models derived from the general model: the 
 unidirectional model (without backward motion, $k_{-}=0$), the random walk (absence of off-pathway pausing state, $k_{p}=0$) and the Poisson model ($k_{-}=$0 and $k_{p}=0$).}
 \label{fig:modelo}
\end{figure*}

The movement of the helicase along DNA can be modelled as a one-dimensional random walk on a chain, where the walker can perform different transitions (Fig.\ref{fig:modelo} (a)): forward movement with probability $P_{+}$ and step $d_{+}$, backward movement with probability $P_{-}$ and step $d_{-}$ and pausing with probability $P_{0}$ and step $d=0$. These probabilities satisfy the normalization condition: $P_{+}+P_{-}+P_{0}=1$. Each transition is governed by an exponentially distributed intrinsic time with an average value: $\tau_{+}$ for the forward transition, $\tau_{-}$ for the backward transition, and $\tau_{0}$ for the pause. This kinetic scheme can be described within the continuous-time random walk (CTRW) framework \cite{wang2020large,kutner2017continuous}. In this formalism, the walker dynamics is described using two stochastic distributions: the jump distribution $f(x)$, which represents the probability of a displacement $x$ during a single step and depends on the probabilities $P_{+}$, $P_{-}$ and $P_{0}$, and the waiting time distribution $\psi(t)$, which describes the time interval $t$ before a transition occurs and depends on the intrinsic times $\tau_{+}$, $\tau_{-}$ and $\tau_{0}$. The probability of locating the walker at a distance $x$ at a time $t$, $P(x, t)$, is given by the Montroll-Weiss expression in Fourier-Laplace space ($s$ is the Laplace transform of $t$ and $k$ the Fourier transform of $x$)\cite{wang2020large,montroll1965random},

\begin{equation}
P(k, s) = \frac{1-\hat{\psi}(s)}{s}\frac{1}{1-\hat{\phi}(k,s)},
\label{Montroll Weiss}
\end{equation}

where $\hat{\psi}(s)$ is the Laplace transform of the waiting time distribution $\psi(t)=\sum_{i=+,-,0}\frac{1}{\tau_{i}}e^{-t/\tau_i}$.
$\hat{\phi}(k,s)$ is the Fourier-Laplace transform of the one-step joint distribution $\phi(x,t)$. By assuming a coupled CTRW where the waiting time and jumps are correlated, 
$\phi(x,t) = \psi(t)f(x|t)$, with $f(x|t)$ representing the conditional probability density function to have a displacement $x$ in  a time interval $t$ during a single step, and the one-step joint distribution reads as:  


\begin{align}
\phi(x,t) = & \frac{P_{-}}{\tau_{-}} e^{-t/\tau_{-}} \delta(x+d_{-}) \notag \\
& + \frac{P_{0}}{\tau_{0}} e^{-t/\tau_{0}} \delta(x) \notag \\
& + \frac{P_{+}}{\tau_{+}} e^{-t/\tau_{+}} \delta(x-d_{+}).
\end{align}

Equation \eqref{Montroll Weiss} can be analytically solved for $k \rightarrow 0$, $s \rightarrow 0$, giving (Supplementary):

\begin{equation}
P(k, s) = \frac{1}{s[1-ik\frac{a_{1}}{sE}-ik\frac{a_{2}E-a_{1}F}{E^2}+\frac{k^2}{2}\frac{c_{1}}{sE}]},
\label{final FL}
\end{equation}

\begin{flalign*}
a_{1}=P_{+}d_{+}-P_{-}d_{-} \\
a_{2}=P_{+}d_{+}\tau_{+}-P_{-}d_{-}\tau_{-}\\
c_{1}=P_{+}d_{+}^2+P_{-}d_{-}^2 \\
E=P_{-}\tau_{-}+P_{0}\tau_{0}+P_{+}\tau_{+}\\
F=P_{-}\tau_{-}^{2}+P_{0}\tau_{0}^{2}+P_{+}\tau_{+}^{2}
\end{flalign*}

Performing the inverse Laplace-Fourier transform we obtain a Gaussian distribution of displacements and times with an average velocity $v$ and a diffusivity $D$ given by:

\begin{equation}
v=\frac{a_{1}}{E}, \quad D = \frac{c_{1}}{2E}+\frac{a_{1}}{2E^2}\left(a_{2}E-a_{1}F\right).
\label{vel and dif supp}
\end{equation}
Details of the model are presented in Supplementary Section V.

\subsection{Data analysis} \label{Data analysis}
The DNA extension as a function of time measured in MT  (Fig.\ref{fig:traces and set up} (b, c, d) upper panels) and OT experiments (Supplementary Fig. S4 (b)) has been converted from nm to bp by using a conversion factor that depends on the force applied. This factor is obtained from the elastic properties of the ssDNA molecule as described in Supplementary Figure S5. 

\begin{figure*}[htbp]
 \centering
 \includegraphics[width=\textwidth]{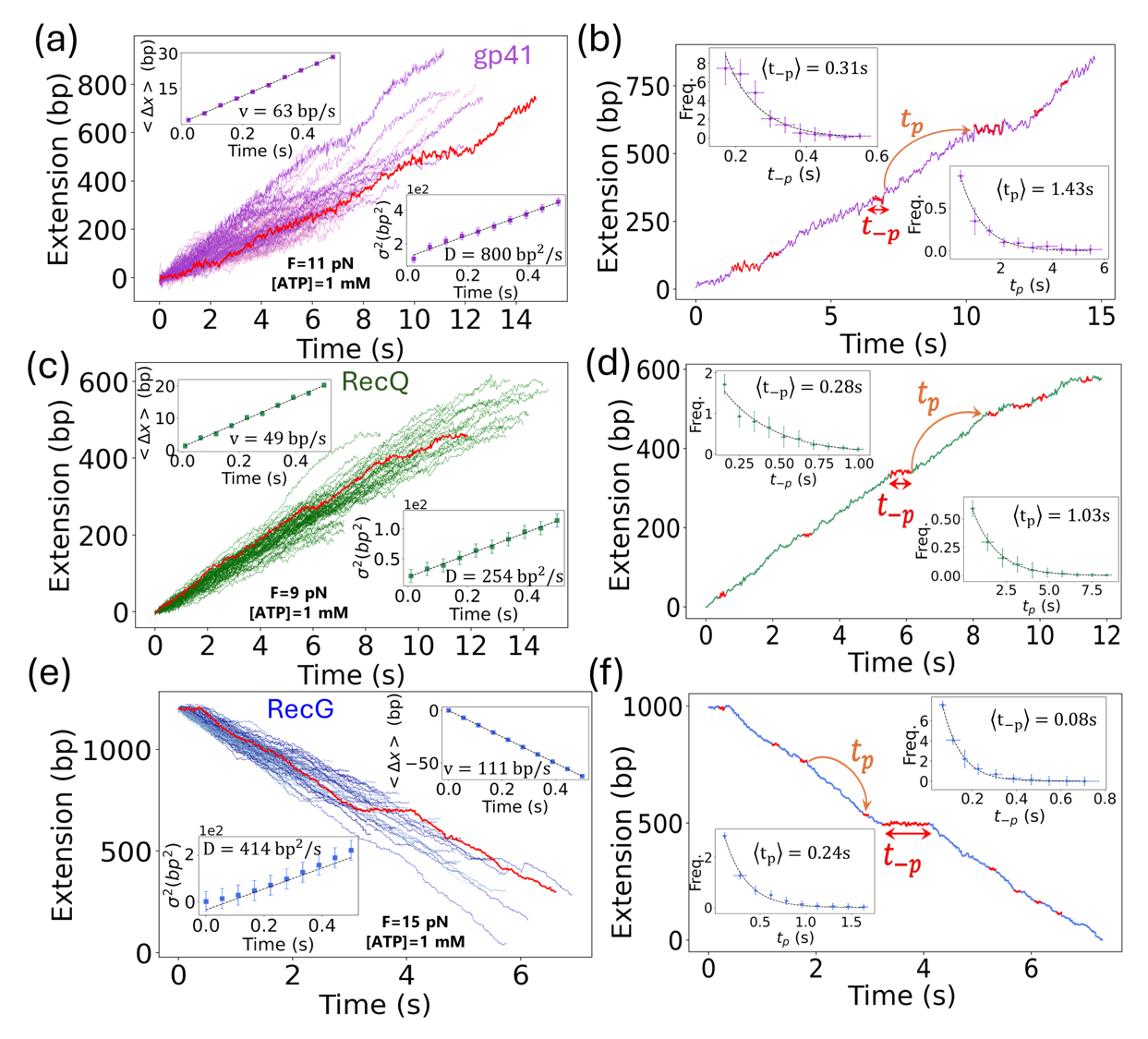}
 \caption{ {\bf Helicase velocity, diffusivity and pause kinetics.} \textbf{(a,c,e)} Set of experimental traces ($\sim$ 50) showing the DNA extension in bps as a function of time for gp41 (purple), RecQ (green) and RecG (blue). A single trace is shown in red as an example. 
 Insets show the mean and variance of the helicase displacement as a function of time, computed from all traces in the main plot. Linear fits are shown as straight lines. Error bars are the standard error of the mean. \textbf{(b,d,f)} A single experimental trace for gp41 (purple), RecQ (green) and RecG (blue), showing the pauses detected with the step-finding algorithm in red. Insets show the distribution of the pause time $t_{-p}$ (top) and the time between pauses $t_{p}$ (bottom) from all the traces in panels (a), (c) and (e). Exponential fits are shown as a continuous lines. Error bars are estimated using the bootstrap method.}
 \label{fig:traces_v_D}
\end{figure*}

In MT experiments, multiple tethers could be tracked simultaneously in a single experiment ($\sim$50), each tether typically exhibiting $\sim$ 10-50 unwinding/rewinding traces. In OT different tethers were tested successively. Typically we used about 3-5 tethers, each presenting $\sim$ 5-20 traces. For each tether (in MT or OT) we computed the average velocity $v$ and the diffusivity $D$. The velocity is determined as the slope of the linear fit of the mean displacement over time, $\langle\Delta x\rangle=vt$ (top insets Fig.  \ref{fig:traces_v_D} (a,c, e)). The diffusivity is determined as half of the slope of the linear fit of the mean square displacement (MSD) over time, $\langle\Delta x^{2}\rangle=2Dt$ (bottom insets Fig.  \ref{fig:traces_v_D} (a,c and e)). The MSD presents an initial regime (at very low times) that deviates from the linear behaviour due to the Brownian motion of the bead (Supplementary Fig. S6) \cite{neuman2005statistical}. We then performed the fits with a time offset of few milliseconds.

We also analysed the pauses along the experimental traces by using a pause detection algorithm based on change point detection \cite{truong2020selective} (Fig. \ref{fig:traces_v_D} (b,d,f) and Supplementary Section VI). The histogram of pausing times can be fitted to an exponential distribution to obtain the characteristic time to exit the pause, $\langle t_{-p} \rangle=\frac{1}{k_{-p}}$ (top insets Fig. \ref{fig:traces_v_D} (b,d and f)). The time lag between pauses also follows an exponential distribution (bottom insets Fig. \ref{fig:traces_v_D} b,d and f ), yielding a value for the characteristic time to enter the pause $\langle t_{p}\rangle=\frac{1}{k_{p}}$.
For each helicase, the rate $v$, diffusivity $D$ and pause kinetics, $k_{-p}$ and ${k_{p}}$, at each experimental condition ([ATP] and force) were computed as the average value between different tethers, typically $\sim$10-50 tethers. 

\section{Results}

To investigate the activity of helicases, we used magnetic and optical traps to mechanically manipulate a DNA hairpin while monitoring the changes in DNA extension as the helicase unwinds or rewinds the hairpin. 
In MT experiments, the DNA hairpin was tethered between a glass surface and a micron-sized magnetic bead, and force was applied using a couple of magnets located on top of the microfluidic chamber (Fig. \ref{fig:traces and set up} (a)). 
In OT experiments, the DNA hairpin was tethered between two micron-sized beads, one held in an optical trap and the other fixed on the tip of a micro-pipette (Supplementary Fig. S4). 
A mechanical force was applied to destabilize the DNA hairpin duplex serving as a means to either assist the advance of a DNA unwinding helicase or hinder the annealing activity of a DNA rewinding helicase. Details of the single-molecule experiments are presented in Section \ref{Single-molecule experiments} in Methods.

\subsection{Measuring DNA unwinding and rewinding activities}

A force above $\sim$15 pN was applied to mechanically unzip the hairpin (Supplementary Fig S2). 
Below $\sim$15 pN, where the hairpin is mechanically stable, we can monitor the DNA unwinding catalysed by helicases that was detected as a smooth increase in the measured extension (Fig. \ref{fig:traces and set up} (b) and (c)). Unwinding activity was measured using MT to track different DNA hairpins in parallel, increasing the statistics.  
Here, we studied two different unwinding helicases: gp41 and RecQ. The gp41 is a hexameric helicase from the T4 bacteriophage that promotes DNA unwinding during DNA replication \cite{lionnet2007real,valentine2001zinc}
, whereas the RecQ is a monomeric helicase from $E.Coli$, playing a central role in DNA repair \cite{bagchi2018single,umezu1990escherichia}.
For studying DNA rewinding helicases, we combined MT and OT experiments, as done in previous works \cite{manosas2013recg}, to explore different force regimes (MT from 5 to 15 pN and OT above 15 pN). In these experiments, we first pulled the extremities of the tethered DNA molecule to partially unzip the hairpin ($\sim$ 15 pN). The enzyme's rewinding (or annealing) activity was detected as a decrease in the DNA extension in MT and OT assays (Fig. \ref{fig:traces and set up} (d) and Supplementary Fig. S3  and Fig S4). Here, we studied RecG, which is a monomeric helicase from $E.Coli$ that promotes the annealing of DNA strands into duplex DNA \cite{manosas2013recg,mcglynn2001rescue}. 
When the annealing activity is coupled to DNA unwinding, it leads to the formation of four-way DNA structures that are crucial in different DNA repair and recombination pathways \cite{singleton2001structural}. 

The measured changes in DNA extension can be converted into a number of unwound/rewound bps from the elastic response of the ssDNA (see Supplementary Fig. S5). This allows to infer the position of the helicase along the DNA (in bp units) as a function of time, as shown in Figure \ref{fig:traces_v_D}. From these traces, we measured the mean and variance of the helicase displacement and extract the helicase velocity and diffusivity (insets in Fig. \ref{fig:traces_v_D} (a,c,e) and Methods). 

\subsection{Measuring helicase ssDNA translocation  activity}
Interestingly, in some cases, we can also monitor the motion of the helicase while it translocates along one strand of ssDNA. In gp41 assays, the experimental traces showed a triangular shape (Fig.\ref{fig:traces and set up} (b) upper panel and Supplementary Fig. S2), where the rising edge corresponds to the helicase unwinding the hairpin, as previously discussed. After the enzyme reaches the loop and the hairpin has been fully unzipped, the helicase can continue translocating on ssDNA while the hairpin re-anneals in its wake. In this latter process, the DNA hairpin annealing reaction, observed as a decrease in the measured DNA extension, is limited by the enzyme translocation. We can then infer the position of the helicase as a function of time in the ssDNA translocation process and estimate the helicase rate and diffusivity, as done with the unwinding traces. For RecQ, the falling edge was not observed (Fig.\ref{fig:traces and set up} (c) upper panel) because RecQ displays strand switching and repeated unwinding when reaching the loop \cite{harami2017shuttling,bagchi2018single}.
This directed motion towards unwinding precluded the detection of the RecQ ssDNA translocation motion. For RecG, we measured its ssDNA translocation activity by performing experiments at low forces using short oligonucleotides to transiently block the DNA fork (see Supplementary Fig. S3 (a) and \cite{manosas2013recg}). After the bound oligonucleotide was displaced, the hairpin's rewinding proceeds at a constant rate, as given by the translocation motion of the helicase. These low-force traces allowed to measure the RecG rate and diffusivity on ssDNA.

\subsection{Model for helicase movement}

Helicase's motion along DNA driven by the nucleotide hydrolysis reaction can be described as a random walk in a one-dimensional chain. Translocation is governed by a set of kinetic reactions that connect different helicase-nucleotide states (or conformations) along the DNA chain. 
The simplest scenario is given by a Poisson model, in which the helicase moves along the DNA in discrete steps of size $d_{+}$ with exponentially distributed waiting times. The average velocity of the enzyme is given by $v=d_+/\tau=d_{+}k_{+}$, where $\tau$ and its inverse $k_+$ are the characteristic waiting time and the forward kinetic rate respectively. The Poisson description is, in general, too simple to capture the dynamics observed in helicases. This is because most helicases exhibit complex mechano-chemical cycles with different rate-limiting steps and multiple pathways. In particular, studies with different helicases have shown the presence of pauses along the helicase trajectories generated by off-pathway states \cite{dumont2006rna, ribeck2010dnab,seol2019homology, craig2022nanopore}. Besides pauses, there are backward steps. In some cases, these backward steps represent intermediate transitions within forward steps \cite{laszlo2022sequence,spies2014two}; in other cases, backward steps reflect slippage events, where the enzyme loses contact with the DNA strand and moves back several bases \cite{manosas2010active, manosas2012collaborative, seol2019homology,schlierf2019hexameric,sun2011atp}. 

Here we propose a minimal CTRW model that incorporates the key features of helicase movement, including forward and backward steps and pauses. In the CTRW model (Fig.\ref{fig:modelo} (a)), transitions are chosen with a probability $P_{+}$ to move right, $P_{-}$ to move left, and $P_{0}$ to enter the pause state with exponentially distributed times of average $\tau_{+}$, $\tau_{-}$ and $\tau_{0}$, respectively. 
For simplicity we assume that forward and backward transitions are characterized by constant steps, $d_{+}$ and $d_{-}$ respectively. 
Using the CTRW framework we can compute the average velocity and diffusivity as a function of the probabilities $P_{+,-,0}$, the transition times $\tau_{+,-,0}$ and the step sizes $d_{+,-}$, Eqs.(\ref{final FL},\ref{vel and dif supp}) (Supplementary Section V and Methods). 
In the context of chemical reactions, kinetic rates $k_i $ are used instead of probabilities $P_i $ and intrinsic transition times $\tau_i $. To express the model in terms of kinetic rates, we consider the following assumptions (see Fig.\ref{fig:modelo} (a)):  
(i) The on-pathway forward reaction and the off-pathway backward (slippage) reaction are irreversible, with rates given by $k_{+} = P_{+}/\tau_{+} $ and $k_{-} = P_{-}/\tau_{-}$, respectively; 
(ii) The off-pathway pause transition is characterized by pause entry and exit rates, given by $k_{p} = P_{0}/\tau_{p}$ and $k_{-p} = 1/\tau_{-p}$, with $\tau_{0}=\tau_{p}+\tau_{-p}$; 
(iii) The intrinsic transition time from the initial state to any other state is assumed to be the same for all transitions $\tau_{+} = \tau_{-} = \tau_{p} = \tau<<\tau_{-p}\rightarrow \tau_{0}\sim\tau_{-p}$ (see Supplementary Section V for details).


The kinetic scheme can be represented as a one-dimensional ladder with two rails (Fig. \ref{fig:modelo} (b)). The lower rail of the ladder stands for the translocation pathway of the ATP-driven forward transition (purple arrow) and the slippage pathway of the backward transition (yellow arrow). The upper rail contains the off-pathway pausing states (red arrows) that are disconnected along the rail.
The average velocity and diffusivity as a function of the kinetic rates is given by (Supplementary Section V and Methods):

\begin{equation}
v=\frac{a_{1}}{E}
\label{eq_vel}
\end{equation}

\begin{equation}
D = \frac{1}{2E} \left[ c_{1} + 2 \left( \frac{a_{1}}{E} \right)^{2} \frac{k_{p}}{k_{-p}} \left( \frac{1}{k_{p}} - \frac{1}{k_{t}} \right) \right]
\label{eq_dif}
\end{equation}

\begin{flalign*}
k_{t} &= k_{+} + k_{-} + k_{-p} \\
E &= \frac{k_{+}}{k_{t}} + \frac{k_{-}}{k_{t}} + \frac{k_{-p}}{k_{p}} \\
a_{1} &= k_{+} d_{+} - k_{-} d_{-} \\
c_{1} &= k_{+} d_{+}^2 + k_{-} d_{-}^2
\end{flalign*}

In general, the different kinetic rates can depend on the concentration of the various reactants (enzyme (E), ATP, ADP, inorganic phosphate $P_{i}$) and the applied force on the experiment. 
As discussed in Supplementary Section VIII, a general expression for the rates $k_i$ is given by: 

\begin{equation}
    \begin{aligned}
        k_{i} &= \frac{k_{cat}^{i}[ATP]}{K_{M}^{i}+[ATP]}e^{\frac{x^{\dagger}_{i}(F-F_{c})}{k_{B}T}}, \quad \text{un(re)winding} \\
        k_{i} &= \frac{k_{cat}^{i}[ATP]}{K_{M}^{i}+[ATP]}, \quad \text{translocation}
    \end{aligned}
    \label{rate dependencies}
\end{equation}
with $i=+, -, p, -p$.
Based on a Bell-like model description \cite{bell1978models}, the rates are exponential with the force $F$ times a transition state distance $x^{\dagger}_{i}$, which is related to the change in DNA extension along the kinetic step $i$. The expression \eqref{rate dependencies} assumes that the force only affects the helicase unwinding/rewinding activity, but not the helicase translocation along ssDNA, where we take $x^{\dagger}_{i}=0$. 
At $F_c\sim 15$pN, where the hairpin mechanically unzips, the helicase rate reduces to the translocation rate that only depends on [ATP]. The  ATP dependence is based on a Michaelis-Menten expression \cite{michaelis1913kinetik} where $k_{cat}^{i}$ is the rate at ATP saturating conditions and $K_{M}^{i}$ is the Michaelis-Menten constant defined as the ATP concentration where the reaction rate is $k_{cat}/2$.  Note that depending on the values of $K_{M}^{i}$ and $k_{cat}^{i}$, the transitions associated to a specific rate $k_i$ would (i) involve ATP hydrolysis (finite $K_M^{i}$ and $k_{cat}^{i}$), (ii) involve ATP binding but not hydrolysis ($K_{M}^{i} \gg [ATP]$), or (iii) not depend on the ATP ($K_{M}^{i} \ll [ATP]$). These different scenarios are explored during the model fitting process (see next section).

\subsection{Best fitting model} \label{subsec:best_fitting_model}

\begin{figure*}[htbp]
 \centering
 \includegraphics[width=0.9\textwidth]{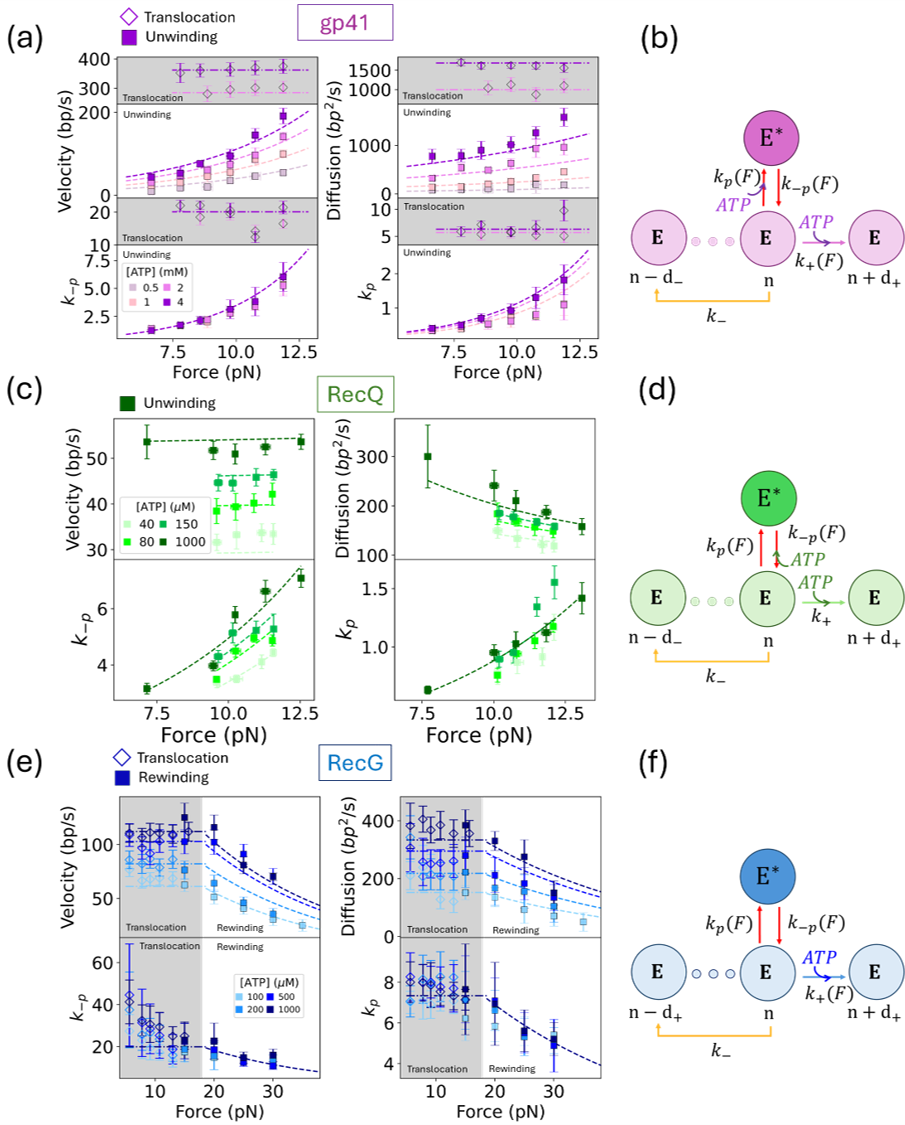}
 \caption{\textbf{Best fitting models} \textbf{(a,c,e)} The measured velocity (top left), diffusivity (top right), exit pause rate (bottom left), and entry pause rate  (bottom right) as a function of force at different ATP conditions for gp41 (purple), RecQ (green), and RecG (blue). The filled squares and empty diamonds are computed from the un(re)winding and translocation data respectively.  Values shown are the mean between different molecules and the error bars represent the standard error of the mean. For gp41 we average $\sim$20 beads with $\sim$20 traces for each bead, for RecQ $\sim$10 beads with $\sim$10 traces each and for RecG $\sim$10 beads with $\sim$10 traces each. The dashed lines are the best fit to the model, as given by Eqs.(\ref{eq_vel},\ref{eq_dif}).
 \textbf{(b,d,f)} Schematic representation of the best-fitting model, showing  the ATP and force dependence of the different kinetic rates. }
 \label{fig:MULTI_FIT}
\end{figure*}

The general model proposed (Fig.\ref{fig:modelo} (b)) considers forward and backward steps of size $d_+$ and $d_-$ and four different kinetic rates, $k_{+}$, $k_{-}$, $k_{p}$, and $k_{-p}$. The rates have their [ATP] and force dependence, as described by Eq.\ref{rate dependencies} , through three independent parameters: $K_{M}$, $k_{cat}$, and $x^{\dagger}$. Summing up the general model has 14 different free parameters. 
The model includes several simplified cases: the Unidirectional model (Uni-model) without backtracking ($k_{-}=0$), the Random walk model (RW-model) without pausing ($k_{p}=0$), and the Poisson model described above ($k_{-}=k_{p}=0$), involving 10, 8, and 4 free parameters, respectively. They are schematically shown in Figure \ref{fig:modelo} (d). 
To reduce the number of free parameters we have analysed the helicase pauses separately. Using a pause detection algorithm \cite{truong2020selective} (Supplementary Fig. S8), we measure the waiting times to enter and exit pauses, $t_{p}$ and $t_{-p}$. Both times follow an exponential distribution (insets in Fig. \ref{fig:traces_v_D} (b,d,f)), from which we derive the average time to enter and to exit the pause, $\langle t_{p} \rangle$ and  $\langle t_{-p} \rangle$,  and the corresponding rates $k_{p}=1/\langle t_{p} \rangle$ and $k_{-p}=1/\langle t_{-p} \rangle$. The exponential behaviour agrees with the single-rate limiting step assumption for a single pause state.
Lower panels of Figure \ref{fig:MULTI_FIT} (a, c, e) show $k_{p}$ and $k_{-p}$ as a function of the applied force for the three enzymes. Rates are exponentially dependent on force as predicted by Eq. \ref{rate dependencies}. The ATP dependence is also well described using the same equation. The overall fit of the force and ATP dependent rates $k_{p}$ and $k_{-p}$ to Eq. \ref{rate dependencies} allows determining the values of $k_{cat}^{p}$, $K_{M}^{p}$, $x^{\dagger}_{p}$, $k_{cat}^{-p}$, $K_{M}^{-p}$ and $x^{\dagger}_{-p}$, reducing the number of model parameters from 14 to 8 for the general model and from 10 to 4 for the Uni-model. The number of fitting parameters for the RW-model and the Poisson model remains unchanged as they do not consider pauses. 

To select the best model that fits the experimental velocity and diffusivity data with the least number of parameters, we performed a least-squares minimization of the reduced chi-square ($\chi^{2}_{\nu}$) value minus 1, where $\nu$ stands for the number of degrees of freedom of the fit, equal to the number of data points minus the number of fitting parameters. The goal is to obtain a value of $\chi^{2}_{\nu}$ as close as possible to 1. Values of $\chi^{2}_{\nu} \gg 1$ indicate a poor model fit, and $\chi^{2}_{\nu} \ll 1$ indicate over-fitting.
We also used the Akaike Information Criterion (AIC) and the Bayesian Information Criterion (BIC) \cite{akaike1974new,schwarz1978estimating}, that are statistical tools for model selection; they give a numerical value that balances the goodness of the fit with the number of parameters (see  Supplementary Section VII).  The optimal model for each helicase was selected as the one that minimizes both AIC and BIC while ensuring $\chi_{\nu}^{2} \approx 1$.
We fitted the general model and the three sub-models (Uni, RW, and Poisson) to the experimentally measured velocity and diffusivity of the three helicases studied: gp41, RecQ, and RecG (Fig. \ref{fig:MULTI_FIT} (a,c,e), upper panels). For each helicase and model, we obtained the AIC and BIC values (Table \ref{tab:tabla_modelos}), selecting the best-fitting model as the model with lowest AIC/BIC values. For the three helicases, the fits to the RW and Poisson models lead to $\chi_{\nu}^{2} >> 1$, showing that these models fail to reproduce the experimental data (Table \ref{tab:tabla_modelos} and Supplementary Fig. S9). In other words, in the absence of an off-pathway pause state, we cannot reproduce simultaneously the measured average rate and diffusivity. 
The best-fitting model for RecQ is the general model.  For gp41 and RecG both the Uni-model and the general model fit the data with similar values of $\chi^{2}_{\nu}$, AIC and BIC (less than 10$\%$ differences in their values, Table \ref{tab:tabla_modelos}). 
Note that the difference between the two models is that the general model includes a backward slippage pathway, whereas the Uni-model does not. Helicase slippage has been previously observed for gp41 \cite{manosas2010active, manosas2012collaborative}, RecQ \cite{seol2019homology} and other helicases \cite{schlierf2019hexameric,sun2011atp}. 
Moreover, large slippage events ($>$ than 10bp) are observed in our experimental traces for the three helicases (Supplementary Fig. S12), but they are not included in the velocity and diffusivity analysis. However, smaller slippage events might be masked in the experimental signal. As a consequence we choose the general model for the three helicases.

Once the model had been selected, we performed a second optimization step for each helicase type by identifying weak dependencies on force and [ATP] of the rates $k_+$, $k_-$, $k_p$, $k_{-p}$ in Eq. \ref{rate dependencies}. We have tested $x^{\dagger}=0$ for the rates that weakly depend on force and $K_M=0$ for the rates that weakly depend on concentration, further reducing the number of parameters. In Table \ref{tab:tabla_modelos} and Figure \ref{fig:MULTI_FIT} we show, for the three helicases, the results from the overall optimization process that leads to the best model with the minimum number of parameters. For gp41 and RecG, we measured the unwinding and ssDNA translocation activities and simultaneously fitted the average rate and diffusivity in the two types of activity, whereas for RecQ we have only used the unwinding data.  Interestingly, for the ssDNA translocation activity, the measured velocity and diffusivity are independent of force for both helicases (Fig. \ref{fig:MULTI_FIT} grey shaded area). This finding supports the view that the main role of the mechanical force is altering the duplex stability, therefore affecting the DNA un(re)winding reaction but not the enzyme translocation along ssDNA.  

For all helicases, the forward rate $k_+$ depends on [ATP], with its Michaelis-Menten constant, $K_M^+$, very close to the $K_M$ obtained by fitting the average un(re)winding rate as a function of [ATP] using the Michaelis-Menten expression:  $v=\frac{k_{cat}[ATP]}{K_{M}+[ATP]}$ (Supplementary Fig. S10 and Table \ref{tab:params}). This shows that the main ATP dependence comes from the on-pathway ATP hydrolysis coupled with the translocating forward step. In contrast, the force dependence is different for each helicase. For RecQ and RecG, $k_{+}$ is weakly dependent on force, whereas for the gp41, $k_{+}$ markedly increases with force (Table \ref{tab:params}), in agreement with the reported active and passive character of these helicases \cite{manosas2010active,manosas2013recg}. On the other hand, the pause kinetics have specific ATP and force dependencies for each helicase. 

The model can be fitted to the experimental data (velocity and diffusivity) with similar values of $\chi^{2}_{\nu} \approx 1$ using a large range of $k_+$ and $d_{+}$ values, which are almost inversely correlated. To limit the range of step sizes, we explored different values around each AIC/BIC minimum and identified a spectrum of values compatible with a relative error less than $5\%$ in both AIC and BIC. This analysis lead to step sizes of $d_{+}=1-3$ bp for gp41, $d_{+}=0.5-1$ bp for RecQ, and $d_{+}=3-4$ bp for RecG. Interestingly, these values are in agreement with previously estimated step sizes for these helicases: $d_{+}=1$ for gp41 or other hexameric helicases \cite{lionnet2007real,schlierf2019hexameric,pandey2014helicase}; $d_{+}=1$ for RecQ \cite{sarlos2012recq,craig2022nanopore} and 
$d_{+}=2-4$ for RecG \cite{martinez2005mechanism,toseland2012atpase, manosas2013recg}. Accordingly, we have chosen $d_{+}=1$ for gp41 and RecQ, and $d_{+}=3$ for RecG. Finally, the backward slippage transition is described with a force and ATP independent rate $k_{-}$ and a backward step $d_{-}\sim 2-6$ bps for the three helicases. 
Recent studies suggest that some helicases might display a variable step size \cite{ma2020dynamic}. 
On the other hand, helicase slippage occurs along a random number of nt and in ATP and force dependent manner (see \cite{manosas2012collaborative} and Supplementary Fig. S12). Therefore the model could be refined by considering variable forward and step sizes $d_{+}$ and $d_{-}$, with force and ATP dependencies.   

\begin{table}[htbp]
\centering
\renewcommand{\arraystretch}{2}
\setlength{\tabcolsep}{4pt}

\begin{tabular}{|c|c|c|c|c|}
\hline
    & $k_+$           & $k_-$           & $k_p$           & $k_{-p}$        \\ \hline

\multirow{2}{*}{\textcolor{purple}{gp41}} & $F$: \textbf{Yes} & $F$: \textbf{No} & $F$: \textbf{Yes} & $F$: \textbf{Yes} \\ \cline{2-5} 
                                  & {[}ATP{]}: \textbf{Yes} & {[}ATP{]}: \textbf{No} & {[}ATP{]}: \textbf{Yes} & {[}ATP{]}: \textbf{No} \\ \hline

\multirow{2}{*}{\textcolor{green}{RecQ}} & $F$: \textbf{No} & $F$: \textbf{No} & $F$: \textbf{Yes} & $F$: \textbf{Yes} \\ \cline{2-5} 
                                  & {[}ATP{]}: \textbf{Yes} & {[}ATP{]}: \textbf{No} & {[}ATP{]}: \textbf{No} & {[}ATP{]}: \textbf{Yes} \\ \hline

\multirow{2}{*}{\textcolor{blue}{RecG}} & $F$: \textbf{Yes} & $F$: \textbf{No} & $F$: \textbf{Yes} & $F$: \textbf{Yes} \\ \cline{2-5} 
                                  & {[}ATP{]}: \textbf{Yes} & {[}ATP{]}: \textbf{No} & {[}ATP{]}: \textbf{No} & {[}ATP{]}: \textbf{No} \\ \hline
\end{tabular}

\caption{Force and ATP dependencies of the rates involved in equation S.17 for the three studied helicases. A schematic diagram is shown in Figure \ref{fig:MULTI_FIT} (b,d,f) with the corresponding dependencies. }
\label{tab:rates}
\end{table}

\begin{table*}[htbp]
\centering
\begin{tabular}{|l||c|c|c||c|c|c||c|c|c|}
\hline
Enzyme & \multicolumn{3}{c||}{gp41} & \multicolumn{3}{c||}{RecQ} & \multicolumn{3}{c|}{RecG} \\
\hline
Estimator & $\chi_{\nu}^2$ & AIC & BIC & $\chi_{\nu}^2$ & AIC & BIC & $\chi_{\nu}^2$ & AIC & BIC \\
\hline
General & \textbf{1.32} & \textbf{970} & \textbf{984} & \textbf{1.53} & \textbf{38} & \textbf{60} & \textbf{1.09} & \textbf{954} & \textbf{969} \\
\hline
Uni & 1.58 & 1041 & 1049 & 10 & 158 & 165 & 1.21 & 1002 & 1014 \\
\hline
RW & 8 & 1309 & 1331 & 9 & 82 & 94 & 5 & 1198 & 1110 \\
\hline
Poisson & 16 & 1493 & 1502 & 11 & 165 & 173 & 27 & 1426 & 1335 \\
\hline
\end{tabular}
\caption{Comparison of models using \(\chi_{\nu}^{2}\), AIC, and BIC. In bold we marked the best values of the estimators.}
\label{tab:tabla_modelos}
\end{table*}

\begin{table*}[htbp]
 \centering
 \begin{tabular}{|l|c|c|c|}
 \hline
 \textbf{Parameter} & \textbf{gp41 ($d_{+}=$1 bp)} & \textbf{RecQ ($d_{+}=$1 bp)} & \textbf{RecG ($d_{+}=$3 bp)} \\
 \hline
 \textbf{$k_{cat}^{+}$} ($s^{-1}$) & 668.52 & 101.13 & 56.51 \\
 \hline
 \textbf{$K_{M}^{+}$ (mM)} & 1.73 & 0.02 & 0.11 \\
 \hline
 \textbf{$x_{+}$ (nm)} & 0.19 & - & -0.04\\
 \hline
 \textbf{$k_{cat}^{-}$} ($s^{-1}$) & 4.22 & 4.98 & 6.11 \\
 \hline
 \textbf{$K_{M}^{-}$ (mM)} & - & - & - \\
 \hline
 \textbf{$x_{-}$ (nm)} & - & - & - \\
 \hline
 \textbf{$k_{cat}^{p}$} ($s^{-1}$) & 7.54 & 1.71 & 7.34 \\
 \hline
 \textbf{$K_{M}^{p}$ (mM)} & 1.02 & - & - \\
 \hline
 \textbf{$x_{p}$ (nm)} & 0.29 & 0.16 & -0.03 \\
 \hline
 \textbf{$k_{cat}^{-p}$} ($s^{-1}$) & 20.13 & 9.15 & 19.94 \\
 \hline
 \textbf{$K_{M}^{-p}$ (mM)} & - & 19.29 & - \\
 \hline
 \textbf{$x_{-p}$ (nm)} & 0.32 & 0.16 & -0.05 \\
 \hline
 \textbf{$d_{-}$} (bp) & 5 & 6 & 3 \\
 \hline
 \end{tabular}
 \caption{Parameters obtained from fitting unwinding and translocation data shown in Figure \ref{fig:MULTI_FIT} using the best fitting model protocol described in section \ref{subsec:best_fitting_model}.}
 \label{tab:params}
\end{table*}

\subsection{On the efficiency of helicases}
The trade-off between the energy cost and the efficiency of helicases can be investigated through the thermodynamic uncertainty relation (TUR) \cite{song2021thermodynamic}. The TUR is an inequality relating the uncertainty (or precision) 
in the motor activity and the energy consumption rate during ATP hydrolysis, known as the entropy production rate $\sigma$.  For an arbitrary current $\dot{x}$ in a nonequilibrium steady state, the time-integrated current $X(t)=\int_0^t \dot{x}(s)ds=x(t)-x(0)$ satisfies the TUR inequality \cite{pietzonka2017finite}:
\begin{equation}
  \sigma\ge \frac{2\langle X(t) \rangle ^2}{V_{X(t)} t}k_B  ,
\label{eq:TUR}
\end{equation}
where $\langle X(t) \rangle$ and $V_{X(t)}=\langle X(t)^2 \rangle-\langle X(t) \rangle^2$ are the mean and variance of the integrated current measured during a time interval $t$. 
From Eq.\eqref{eq:TUR} one can define the dimensionless $Q$ factor that quantifies the tightness of the TUR inequality \cite{song2021thermodynamic},

\begin{equation}
  Q=\frac{\sigma V_{X(t)}}{k_B\langle X(t) \rangle ^2}t\ge 2. 
\end{equation}

For helicases, $X$ is the motor displacement measured by the bead's position. From the velocity $v=\frac{\langle X(t) \rangle}{t}$ and diffusivity $D=\frac{V_{X(t)}}{2t}$ we get, 
\begin{equation}
  Q=\sigma\frac{2D}{k_{B}v^2}\ge 2.
\label{eq:Q factor}
\end{equation}

In general, for translocating motors, the TUR bound is loose with $Q$ values ranging from 5 to 20 for kinesin, from 5 to 13 for myosin, or from 50 to 100 for T7 DNA polymerase and the ribosome \cite{hwang2018energetic,song2020thermodynamic,pineros2020kinetic}. In previous studies \cite{song2021thermodynamic}, molecular motors have been grouped within a similar range of $Q$ values. However, to our knowledge, $Q$ has never been investigated for helicases. 
We determine $v$ and $D$ from the time traces of the motor, but $\sigma$ is not directly measured. However, $\sigma$ can be estimated assuming a tight mechano-chemical coupling between the unwinding-rewinding of $d_{+}$ bps and the hydrolysis of one ATP \cite{sun2011atp,xie2020non}. Moreover, we assume that ATP is not consumed in the backward and pausing steps. If $\Delta G$ is the average (positive) energy consumed in one step, the consumed energy rate $\sigma$ can be estimated as: $\sigma = \frac{\Delta G}{T} \frac{v}{d_{+}}$ with $d_{+} = 1$ bp for gp41 and RecQ and $d_{+} = 3$ for RecG. The energy balance has chemical and mechanical contributions, $\Delta G = \Delta\mu-W$, where $W=d_+(\Delta G_{\rm bp}+W_F)$ is the reversible mechanical work needed to unzip $d_+$ bp at force $F$. $\Delta G_{bp}$ is the (positive) hybridization free energy per bp, $\Delta G_{bp} \approx 2 k_{B}T$ \cite{huguet2010single}. The stretching contribution at force $F$ equals $W_{F}=-2\int_{0}^{F} x_{WLC}(f ^{\prime}) df ^{\prime}$ where $x_{WLC}(f ^{\prime})$ is the extension per base at force $f'$ and we use the worm-like chain (WLC) model for ssDNA elasticity \cite{mossa2009dynamic}. $W_{F}$ changes from 0 to $-2 k_{B}$T between $F=0$ and the coexistence force $F_{c}\sim 15pN$ (Supplementary Fig. S11). Note that at $F_{c}$, $W_F=-\Delta G_{\rm bp}$ and $W=0$, meaning that the energy from ATP hydrolysis is fully dissipated as heat. For a rewinding helicase, both $\Delta G_{bp}$ and $W_{F}$ change sign: $\Delta G_{bp}=-2 k_{B}T$ is the average energy to melt one bp and $W_{F}=2\int_{0}^{F} x_{WLC}(f ^{\prime}) df ^{\prime}$ corresponds to the work required to bring the two bases from force $F$ to zero force.   
Finally, the energy released from ATP hydrolysis $\Delta \mu \approx 12-20 k_{B}T$, depending on the ATP concentration. 

The $Q$ factor is related to the thermodynamic efficiency. The second law implies $\sigma\ge 0$, and therefore $\Delta G\ge 0$ or  $W\le \Delta \mu$. We define the motor efficiency $\eta$ as the ratio between the amount of mechanical work per step $W=d_+\cdot(\Delta G_{\rm bp}-W_F)$ and the available chemical energy from ATP hydrolysis $\Delta\mu$, $ \eta=\frac{W}{\Delta\mu}$. Using the energy balance, $\Delta G = \Delta\mu-W$ we can write the efficiency as,
\begin{equation}
    \eta=\frac{W}{\Delta\mu}=\frac{1}{1+\Delta G/W}=\frac{1}{1+\frac{\sigma T d_{+}}{vW}},
    \label{eff}
\end{equation}

For the case of a molecular motor un(re)winding DNA Eq. \eqref{eq:Q factor}, $\eta$ can be written as \cite{pietzonka2016universal,song2021thermodynamic}:

\begin{equation}
    \eta=\frac{1}{1+Q\frac{vd_{+}}{2D}\frac{k_{B}T}{W}}.
\end{equation}

\begin{figure}[htbp]
 \centering
 
\includegraphics[width=\columnwidth]{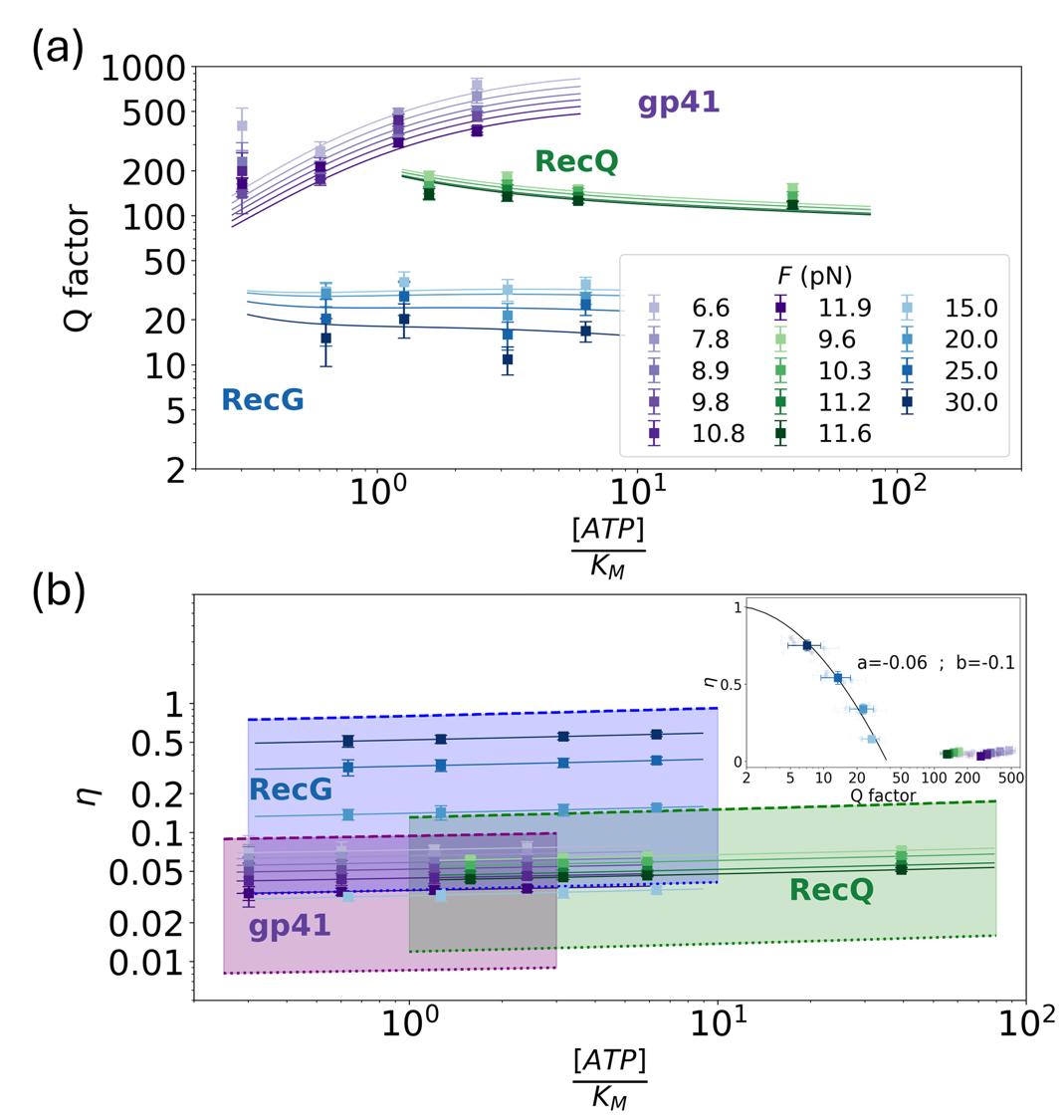}
 \caption{ (a) Estimated $Q$ factor as a function of ATP concentration for different forces in a log-log scale for the three studied helicases, gp41 in purple, RecQ in green and RecG in blue. The ATP concentration is divided by the Michaelis-Menten constant of each helicase. (b) Efficiency $\eta$ as a function of the ATP concentration for different forces in a log-log scale. The inset shows $\eta$ as a function of $Q$ for the three helicases in a linear-log scale. The continuous line shows the fit to the RecG data of the function  $\eta=1+a\log(Q/2)+b\log^2(Q/2)$ with $a=-0.06, \ b=-0.1$.}
 \label{fig:Q_factor}
\end{figure}

To calculate $Q$ and $\eta$ we use Eqs.(\ref{eq:Q factor},\ref{eff}) using the measured values of $v,D$ and the estimated $\sigma$.
In Figure \ref{fig:Q_factor} we show $Q$ and $\eta$ for the three studied helicases in a log-log scale, as a function of the ATP concentration at different forces.
RecQ and gp41 present large Q values $100-200$ and $200-500$, respectively, and low efficiencies at zero force around 0.1, which further decrease with force. In contrast, RecG has lower Q values $\sim 10-20$, which decrease upon increasing the force above $F_c$. Interestingly, the lower $Q$ values correlate with a higher $\eta$, with RecG reaching an efficiency close to 1 at the stalling force of $\sim 35-40$pN \cite{manosas2013recg}. In the inset of Fig \ref{fig:Q_factor} (b), we plot $\eta$ versus $Q$ in a linear-log scale. While gp41 and RecQ fall at the bottom right inefficiency corner of high Q-low $\eta$ values, RecG follows a trend with $\eta$ increasing upon decreasing $Q$, approaching its maximum $\eta=1$ if $Q\to 2$. A two-parameter fit to the function $\eta=1+a\log(Q/2)+b\log^2(Q/2)$ gives $a=-0.06, \ b=-0.1$ (inset, continuous black line). The significance of this fit lies in the logarithmic dependence of $\eta$ with $Q$, underlying a fundamental looseness of the TUR regarding the thermodynamic efficiency of molecular machines.  
Comparing the unwinding helicases, RecQ, and gp41, we observe that  $Q$ depends on the passive and active nature of the enzymes with larger $Q$ values for passive helicases, yet $\eta$ remains qualitatively similar.


\section{Discussion and conclusions}

DNA replication and repair are fundamental processes of life by which genetic information is preserved and transferred to the next generation. These processes require the action of helicases that use ATP hydrolysis to move on DNA inducing the unwinding and rewinding of the double helix \cite{lohman1996mechanisms,tuteja2004unraveling,delagoutte2003helicase,pyle2008translocation}. Single-molecule techniques monitor the activity of individual helicases, revealing a non-monotonic rate with forward steps, backward steps, and pauses \cite{dumont2006rna,ribeck2010dnab,spies2014two,seol2019homology,laszlo2022sequence}. The complexity of the mechanochemical reactions at the nanoscale in a highly noisy environment makes their study challenging. 

In this work, we use magnetic and optical traps to monitor the motion of different DNA helicases (gp41, RecQ and RecG), while they move along DNA under different forces and ATP concentrations (Fig. \ref{fig:traces and set up}). 
The helicase motion can be characterized by measuring the velocity when the enzyme catalyzes the unwinding or rewinding reaction and the velocity when it translocates along a single DNA strand. As shown in previous works, an analysis of how these velocities depend on force and ATP concentration gives information about the helicase mechano-chemical cycle \cite{manosas2010active,manosas2013recg,lionnet2007real,ribeck2010dnab,spies2014two,seol2019homology,laszlo2022sequence}. Here we extend the analysis by measuring the helicase diffusivity, which allows us to better characterize the helicase dynamics (Fig.\ref{fig:traces_v_D}). We use a Continuous Time Random Walk (CTRW) model,  depicted in Figure \ref{fig:modelo}, to describe the helicase dynamics that include forward and backward steps and pauses. The CTRW formalism presented enables us to obtain analytical expressions for the helicase velocity and diffusivity, given in Eqs.(\ref{eq_vel},\ref{eq_dif}). Fitting the model to the data allows us to determine the minimum number of states and transitions for each helicase (Fig.\ref{fig:MULTI_FIT}). 
Introducing an off-pathway pause state into the CTRW model is essential to describe the velocity and diffusivity for the three helicases studied. Models without pauses, such as the Poisson or Random Walk model, do not fit the data, raising the question of the biological role of pauses. Helicases work in coordination with other enzymes to perform their biological functions. The RecQ and RecG helicases work with single-stranded binding proteins and other accessory proteins in different DNA repair pathways. Gp41 operates as part of a large complex containing two polymerases and other proteins, known as replisome and responsible for replicating the genomic DNA in T4 bacteriophage. Previously, we have shown that when the gp41 helicase works together with the polymerase, the rate of the helicase advance increases without pausing \cite{manosas2012collaborative}. 
Therefore, pausing might be the strategy to control helicase activity. Without the accessory proteins needed to develop a specific biological function, such as replication and repair, pauses stall the helicase activity. 

A general feature observed for all helicases is that force affects unwinding and rewinding activity but not the translocation activity along one strand of DNA (Fig.\ref{fig:MULTI_FIT} (a,e), white versus gray background).
The main effect of force is to destabilize the DNA duplex in a helicase-dependent manner. For gp41, the velocity and diffusivity are very sensitive to the value of the applied force, whereas for RecQ and RecG, they are not. Indeed, when the force changes by 5 pN, the rate and diffusivity change by a factor of ten for gp41, whereas they remain almost constant for RecQ and RecG (Fig.\ref{fig:MULTI_FIT} a,c,e). This force sensitivity is related to their active and passive character, as discussed elsewhere \cite{manosas2010active}.

Measuring velocity and diffusivity is also useful to study the thermodynamic efficiency $\eta$ and the thermodynamic uncertainty relation (TUR) through the associated $Q\ge 2$ value, Eq.\eqref{eq:Q factor}. 
RecG presents the smallest Q factor and largest efficiency as compared to gp41 and RecQ (Fig.\ref{fig:Q_factor}). In particular, at forces close to the stalling force $\sim 40$ pN, RecG reaches $\eta\sim$1 by operating close to the thermodynamic optimization limit, $Q=2$. This large efficiency correlates with its large step size of 3 bp. In contrast, gp41 and RecQ, which unwind only one bp per ATP hydrolysed, present much lower $\eta$, below 0.15. Remarkably, only RecG during rewinding shows efficiencies approaching 1, whereas unwinding activity for gp41 and RecQ helicases is thermodynamic inefficient, with most of the energy from ATP hydrolysis released as heat. This fact indicates that the efficiency of molecular machines is largest whenever they operate uphill in the sense that the energy cost of the task $W$ is comparable to the chemical energy from ATP hydrolysis, $\Delta\mu$. A similar phenomenon occurs for ATPase \cite{yasuda1998f1}, which is almost 100$\%$ efficient when transporting protons against the electrochemical potential gradient to synthesize ATP. In contrast, in the presence of thermogenic proteins, ATP synthase decouples from the proton gradient, and the proton flow is employed to produce heat.  

The CTRW framework proposed in this work allows for characterizing helicase dynamics through velocity and diffusivity measurements, which can be easily obtained from single-molecule assays. The model can be easily adapted to describe motors that translocate along DNA or through other templates, such as polymerases or kinesins.

\section{Acknowledgements}

Authors thank P. Bianco for RecG protein proparations. V.R-F, M.M. and F.R. acknowledge support from
the Spanish Research Council Grant [PID2019-111148GB-100]; M.M. acknowledge support from MICIU/AEI/10.13039/501100011033 and NextGenerationEU/PRTR [CNS2022-135910]. F. R. also acknowledges support from ICREA Academia Prizes 2018 and 2023.

\section{Conflict of interest statement}
None declared.

\bibliographystyle{unsrt}
\bibliography{bibliography}

\end{document}